\documentclass[aps,prb,twocolumn,showpacs,amsmath,amssymb,floatfix]{revtex4-1}
\usepackage{graphicx}
\usepackage{dcolumn}
\usepackage{bm}

\begin{document}

\title{Phase transitions in single-crystalline magnetoelectric LiCoPO$_{4}$}

\author{A. Szewczyk}
\email{szewc@ifpan.edu.pl}

\author{M. U. Gutowska}

\author{J. Wieckowski}

\author{A. Wisniewski}

\author{R. Puzniak}

\author{R. Diduszko}

\affiliation{Institute of Physics, Polish Academy of Sciences, Al. Lotnikow 32/46, 02-668 Warsaw, Poland}

\author{Yu. Kharchenko}

\author{M. F. Kharchenko}

\affiliation{B. Verkin Institute for Low Temperature Physics and Engineering, National Academy of Sciences of Ukraine, pr. Lenina 47, 61103 Kharkiv, Ukraine}

\author{H. Schmid}
\affiliation{Department of Inorganic, Analytical and Applied Chemistry, University of Geneva, 30 quai Ernest-Ansermet, 1211 Geneva 4, Switzerland}

\begin{abstract}

 Specific heat, magnetic torque, and magnetization studies of LiCoPO$_{4}$ olivine are presented. They show that an unique set of physical properties of LiCoPO$_{4}$  leads to the appearance of features characteristic of 2D Ising systems near the N\'{e}el temperature, $T_N$\,=\,$21.6$ K, and to the appearance of an uncommon effect of influence of magnetic field on the magnetocrystalline anisotropy. The latter effect manifests itself as a first-order transition, discovered at $\sim$\,9 K, induced by magnetic field of 8 T. Physical nature of this transition was explained and a model describing experimental dependences satisfactorily was proposed.

\end{abstract}

\maketitle

\section*{introduction}

LiCoPO$_{4}$ olivine, crystallizing in the \textit{Pnma} structure,\cite{santo} Fig.\ \ref{fig1}, exhibits a unique set of physical properties, which makes it attractive for both basic and applied studies. That means:
\begin{enumerate}
\item[(i)] It shows an exceptionally large linear magnetoelectric effect\cite{rive, korn01, korn02} and a large Li-ionic conductivity (making it promising for application as cathodes in Li-ions batteries).\cite{han, chung, molen}
\item[(ii)]In its structure, (100) oriented, ``corrugated" Co-O layers can be distinguished, within which the Co$^{2+}$ magnetic moments are strongly coupled by superexchange Co--O--Co interactions. The neighboring (100) layers are coupled weakly by higher order interactions,\cite{vak, mays} e.g., Co--O--P--O--Co. Below $T_N$\,=\,$21.6$ K, an antiferromagnetic ordering appears in the system. Due to large anisotropy, \cite{vak, tian} Co magnetic moments are confined to the directions lying within the \textit{b--c} plane, ca. 4.6$^{\circ}$ away from the \textit{b} axis. The magnetoelectric effect studies\cite{rive} (revealing ``butterfly" hysteresis loops and a possibility to produce the single domain state by application of a magnetic field alone), as well as the direct magnetization measurements \cite{khar27, khar28} showed that the Co magnetic moments do not compensate each other completely and a small net magnetic moment, parallel to the \textit{b} axis, is present. Thus, LiCoPO$_{4}$ is an intriguing quasi - two-dimensional weakly ferromagnetic Ising system.
\begin{figure}
\includegraphics[scale=0.8]{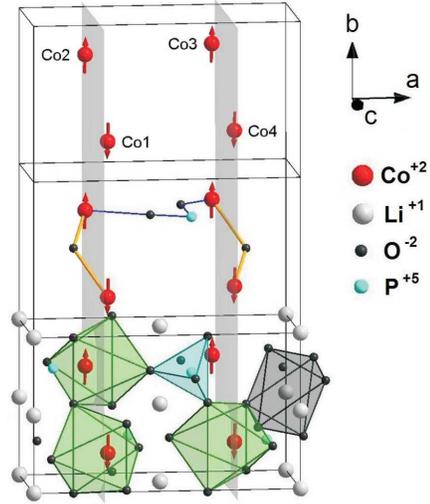}
\caption{\label{fig1}(Color online). Orthorhombic (\textit{Pnma}) olivine structure of LiCoPO$_4$. Three unit cells ($a$\,=\,10.20 \AA, $b$\,=\,5.92 \AA , $c$\,=\,4.70 \AA) stacked along the \textit{b} axis are presented. Starting from the lowest one, they show, respectively, oxygen coordinations of Co (octahedral), Li (octahedral) and P (tetrahedral) ions, the examples of strong Co--O--Co and weak Co--O--P--O--Co superexchange couplings, and the antiferromagnetic ordering of magnetic moments of the Co ions.}
\end{figure}
\begin{figure*}
\includegraphics[scale=0.65]{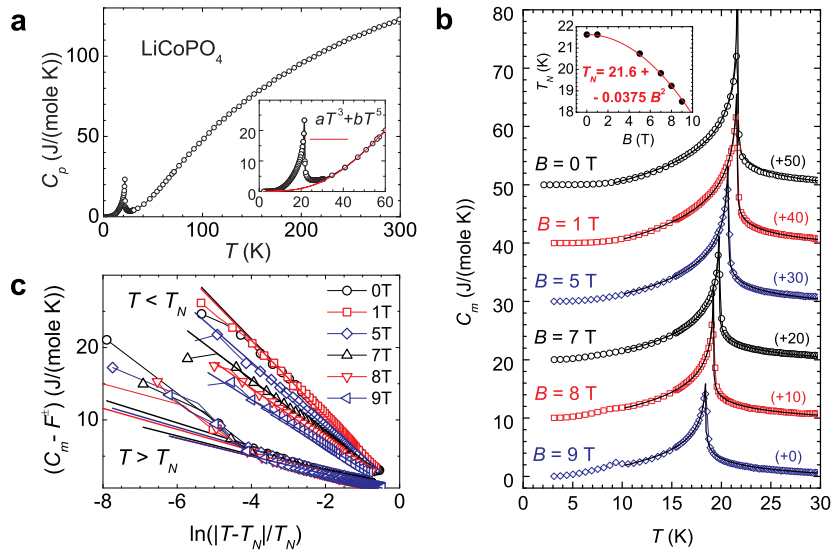}
\caption{\label{fig2}(Color online).
Specific heat of LiCoPO$_{4}$. (a) Temperature dependence of the total specific heat in zero magnetic field, $B$. Inset shows the $\lambda$-anomaly near the N\'{e}el temperature, $T_N$. The parameters a $= 1.31 \times 10^{-4}$ J/(mole K$^4$) and b $= -1.06 \times 10^{-8}$ J/ (mole K$^6$) determine lattice contribution. (b) Magnetic contribution to the specific heat, $C_m$, as a function of temperature, measured on heating, in $\mathbf{B}$ parallel to the \textit{b} axis. Curves for different $B$ values are shifted along the $C_m$ axis by the values given in parentheses. Inset shows the dependence of $T_N$ on $B$ (experimental points and fitted parabola). The solid lines present logarithmic dependences fitted to the experimental data near $T_N$. (c) $C_m - F^{\pm}$ vs. $\ln(\tau)$, $\tau=|T-T_N|/T_N$. Solid lines are linear approximations valid for $ \sim e^{-5}<\tau<\sim e^{-0.6} $ above and below $T_N$.}
\end{figure*}
\item[(iii)] The presence of a spontaneous magnetization is not consistent with the \textit{Pnma'} symmetry, for years assumed to be the magnetic symmetry of LiCoPO$_{4}$, but is consistent with the $P12_1^{'}1$ monoclinic symmetry, in which, additionally, a nonzero dielectric polarization and a nonzero toroidal moment are allowed (the monoclinic \textit{b} axis coincides with the pseudo-orthorhombic \textit{a} axis). Attempts at measuring the dielectric polarization were unsuccessful.\cite{aken} A nonzero toroidal moment was derived \cite{aken, eder} on the microscopic level, based on the magnetic structure data.\cite{vak} On the macroscopic level, four domain states were observed,\cite{aken} two of which were interpreted as "antiferromagnetic" and two other ones as "ferrotoroidic". However, detailed symmetry considerations\cite{schmid} showed that all four domain states are equivalent and differ in orientation of the net magnetic moment. Each of the domains bears a net magnetic and a toroidal moment, whose signs and directions are mutually rigidly coupled.
\item [(iv)] The studies of birefringence induced by magnetic field \cite{khar2000} suggest that the magnetic structure can be even more complex. In addition to the large, uniform in space, and parallel to the \textit{b} axis component of the main antiferomagnetic vector, $\mathbf{L}_2=\mathbf{m}_1-\mathbf{m}_2-\mathbf{m}_3+\mathbf{m}_4$ ($\mathbf{m}_i$ are magnetic moments of Co ions), small, modulated in space, perpendicular to the \textit{b} axis components of $\mathbf{L}_2$ and of other antiferromagnetic vectors, defined in Ref. \onlinecite{khar28}, can exist.
\end{enumerate}

Despite intensive studies of  structural,\cite{santo} magnetic,\cite{santo, khar28, khar27, vak} magnetoelectric,\cite{rive, korn01, wieg} transport,\cite{wolf} and optical\cite{aken, khar2000, korn99, fom, khar08} properties, actual magnetic and electric structures of LiCoPO$_{4}$ and their transformations in magnetic field have not yet been elucidated satisfactorily.

Since specific heat is very sensitive to all phase transitions, this work was aimed at studying thermal properties of LiCoPO$_4$, at determining the order of observed  phase transitions (spontaneous and induced by magnetic field, $\mathbf{B}$, applied along the \textit{b} axis) and at investigating how the intermediate dimensionality of the magnetic structure of LiCoPO$_4$ influences the critical behavior near $T_N$.

\section*{experiment}

For the present studies, a LiCoPO$_4$ single crystal obtained by high temperature solution growth using lithium chloride (LiCl) as flux was chosen. In Ref. \onlinecite{merc}, this method of crystal growth was shown to be applicable for the entire crystal family LiMPO${_4}$ (M= Ni, Co, Fe, Mn). In the present case, the synthesis was realized in full analogy to that described in detail for LiNiPO${_4}$ in Ref. \onlinecite{fom02}, i.e., using a molar ratio 1:3 between LiCoPO$_4$ and LiCl in the starting mixture and using sealed platinum crucibles with 30 ml volume, with a 50 $\mu$m hole in the lid for equilibrating the pressure and minimizing loss of the highly volatile LiCl solvent. The growth parameters and the special technique for separating the flux from the crystals were identical with those used for LiNiPO${_4}$. \cite{fom02} No impurity phases occur in the described synthesis process. The growth morphology of the LiCoPO$_4$ crystals has been described in detail in Ref. \onlinecite{rive} and is characterized by the development of orthorhombic (100), (210), (011), and (101) facets, which may be used as reference for the preparation of samples (even without X-ray orientation).

The specific heat of the LiCoPO$_4$ single crystal was measured by means of the relaxation method, using the Physical Property Measurement System, PPMS, made by Quantum Design. Estimated uncertainty of the determined specific heat values was $\sim 2\%$ . In $B$\,=\,0, studies were done from 2 to 300 K. Since no phase transitions appeared above $T_N$, temperature dependences for nonzero $B$ values, ranging from 1 up to 9 T, were measured up to 40 K only (the magnetic field, $\mathbf{B}$, was applied along the \textit{b} axis). The experimental points were measured every 0.3 K (for $B$\,$\neq$\,$0$) or 0.2 K (for $B$\,=\,0) below 15 K and every 0.1 K within the critical region around $T_N$. In all figures, not all experimental points are marked with symbols to keep legibility.
Supplementary magnetization and magnetic torque measurements have been performed by using respective measurement options of PPMS.

\section*{results}

\begin{figure}
\includegraphics[scale=1]{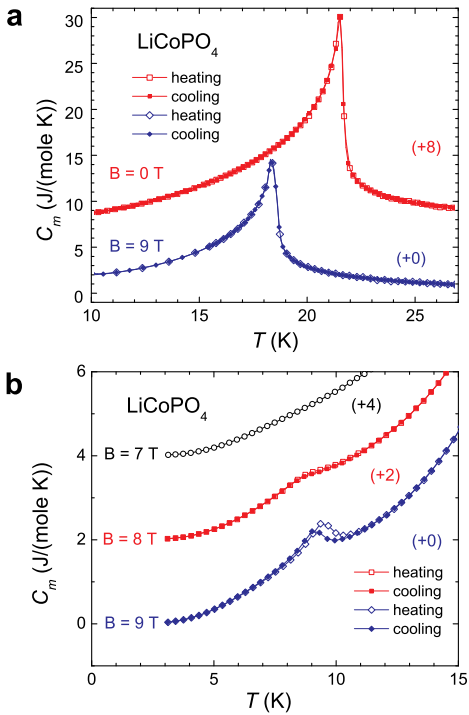}
\caption{\label{fig3}(Color online).
Magnetic contribution to the specific heat of LiCoPO$_4$, $C_m$, measured on heating and on cooling. Curves for different magnetic field values are shifted along the $C_m$ axis by the values given in parentheses. (a) Lack of hysteresis around $T_N$. (b) Thermal hysteresis near $T=9$ K for $B=9$ T.}
\end{figure}

The zero-field temperature dependence of specific heat is plotted in Fig.\ \ref{fig2}a.\ The inset shows the $\lambda$-shaped anomaly accompanying the paramagnetic - weakly ferromagnetic phase transition at $T_N$. Due to low electric conductivity,\cite{wolf} $\sim$\,$10^{-9}$ Scm$^{-1}$, the electronic contribution to the specific heat is negligible and the total specific heat,  $C_p$, consists of the lattice, $C_\text{ph}$, and magnetic, $C_m$, contributions only. Below 60 K, $C_\text{ph}$ can be described by the formula:
 \begin{equation}
 C_\text{ph}\left (T\right )=\text{a}T^3+\text{b}T^5, \qquad   \text{a}=7N_Ak_B \frac{12\pi^4}{5\theta^3_{D}},
\label{eq1}
\end{equation}
where the term $\sim T^3$ represents the low-temperature dependence in the Debye model and the term $\sim T^5$ is the correction of that model \cite{grim} for the nonlinear phonon dispersion relation:
$\omega=\text{c}_1 \vert k \vert + \text{c}_2 \vert k \vert^2$. Based on Eq.\ (\ref{eq1}), the Debye temperature was estimated to be $\theta_D=470$ K.

\begin{table}
\caption{\label{tab1}Parameters fitting the logarithmic dependence, Eq.\ (\ref{eq2}), to the experimental data the best. A$^{\pm}$ and F$^{\pm}$ are given in J/(mole K). For each parameter, an estimated uncertainty of the last digit (or of the two last digits) is given in parentheses, e.g., 21.63(3) means $21.63\pm 0.03$.}
\begin{ruledtabular}
\begin{tabular}{ccllll}
$B$(T) & $T_N$(K) &  ~A$^-$ & ~~F$^-$ &  ~A$^+$ & ~~F$^+$\\
\hline
0&21.63(3)&5.3(2)& \text{$-2.6(3)$}&1.6(1)&$-0.9(5)$\\
1&21.63(1)&5.25(5)& \text{$-2.6(2)$}&1.45(5)&$-0.8(2)$\\
5&20.74(3)&4.5(2) & \text{$-1.75(10)$}&1.5(1)&$-0.8(1)$\\
7&19.82(2)&3.9(1)& \text{$-1.2(1)$}&1.3(1)&$-0.35(10)$\\
8&19.23(5)&3.6(1)& \text{$-1.0(1)$}&1.25(10)&$-0.25(10)$\\
9&18.50(1)&3.2(1)& \text{$-0.67(10)$}&1.25(5)&$-0.05(10)$\\
\end{tabular}
\end{ruledtabular}
\end{table}

The magnetic contributions to the specific heat, determined by subtracting the $C_\text{ph}\left (T\right )$ calculated according to Eq.\ (\ref{eq1}) from the measured specific heat, as well as evolution of $T_N$ with $B$ are presented in Fig.\ \ref{fig2}b. To determine the order of the phase transition occurring at $T_N$, $C_p\left (T\right )$ for $B$\,=\,0 and 9 T was measured on heating and on cooling. For both field values, no thermal hysteresis was detected, Fig.\ \ref{fig3}a, which strongly suggests that this is a second order transition and that its order does not change in magnetic field. It was verified that the experimental data can not be described appropriately by assuming the classical form \cite{bin} of critical behavior:
$C_m\sim \left (\vert T - T_N \vert / T_N \right )^{-\alpha}$, where the critical exponent $\alpha$ takes a value between 0 (corresponding to the logarithmic divergence for the two-dimensional, 2D, Ising system \cite{onsag}) and $\sim$\,0.119 (found for the three-dimensional Ising model\cite{bin}). Thus, the critical behaviour in the form: \cite{woodf}
\begin{equation}
C_m(T) = \begin{cases}
          - \text{A}^+\ln \left (\frac{T-T_N}{T_N} \right ) + \text{F}^+  & \text{for \; $T>T_N$}\\
              -\text{A}^-\ln  \left (\left | \frac{T-T_N}{T_N} \right  | \right ) + \text{F}^-   &\text{for \; $T < T_N$}
           \end{cases}
\label{eq2}
\end{equation}

\begin{figure*}
\includegraphics[scale=0.9]{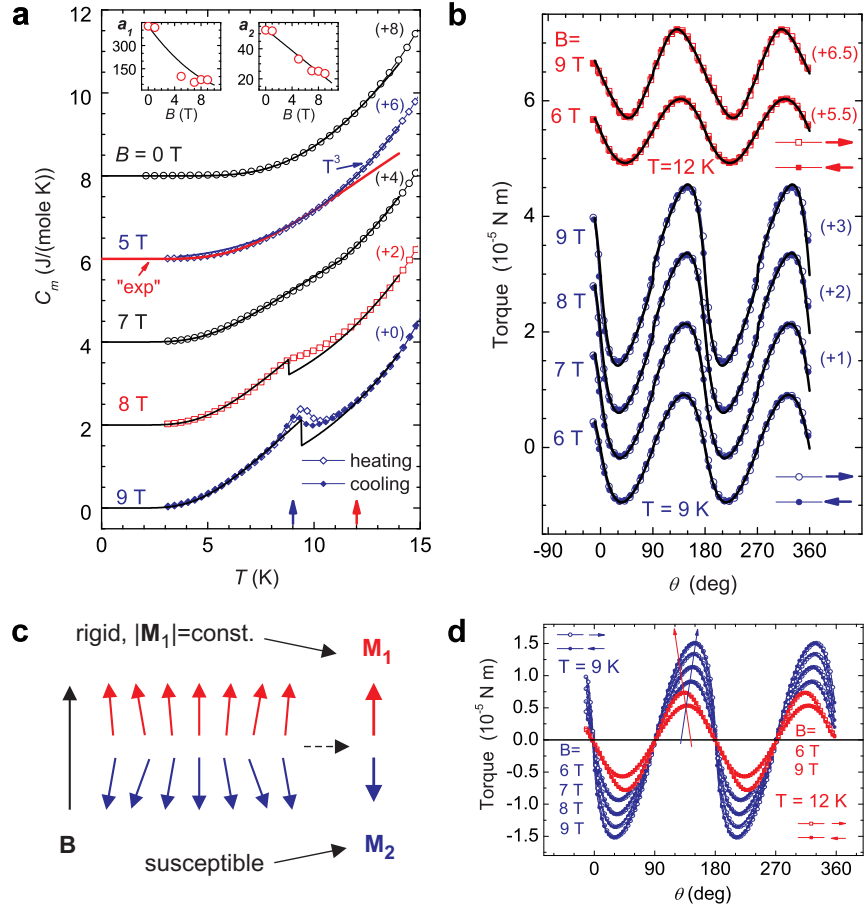}
\caption{\label{fig4}(Color online).
Field-induced first-order phase transition. Curves in panels (a) and (b) are shifted along the \textit{y}-axes by the values given in parentheses. (a) Magnetic specific heat vs. temperature. For $B \geqslant 8$ T, the anomaly at $\sim$\,9 K is visible. Vertical blue and red arrows indicate the temperatures, at which torque was measured. Solid lines present the calculated magnon contributions. The insets show experimental (circles) and fitted theoretical (solid lines) dependences of two parameters appearing in Eq.\ (\ref{eq3}), $a_1$ (in J/(mole K$^{1/2}$)) and $a_2$ (in K), on $B$. (b) Magnetic torque for $\mathbf{B}$ rotating within the \textit{b-c} plane ($\theta $ is counted from the \textit{b} axis). Curves plotted with full and open symbols were measured for $\mathbf{B}$ rotating in opposite directions. Dependences calculated within the proposed model (solid lines) are superimposed on the experimental curves. (c) Outline of the proposed two-sublattice model of the magnetic structure. (d) Superimposed magnetic torque dependences measured at $T = 9$ and 12 K. The oblique arrows indicate directions along which the torque maximum moves with increasing $B$ at $T = 9$\ K and at $T = 12$\ K.}
\end{figure*}
was assumed and the satisfactory description of the experimental data has been achieved, Figs.\ \ref{fig2}b and \ref{fig2}c. On the contrary to the ideal 2D Ising system,\cite{onsag} the anomaly at $T_N$ is evidently asymmetric with respect to $T_N$, Fig.\ \ref{fig3}a, which suggests that both A and F parameters have different values for both sides of $T_N$. This qualitative expectation was confirmed by calculations and the A$^\pm$, F$^\pm$, and $T_N$ values fitting the experimental data the best are given in Table \ref{tab1}. (It should be stressed that the uncertainty of $T_N$ given in Table \ref{tab1} is the uncertainty of the theoretical, fitted parameter. For temperatures $\sim 30$ K, the uncertainty of the absolute temperature values determined in PPMS is $\pm 1\% $, whereas relative temperature changes $\sim 0.03\%$ can be detected an stabilized). We attribute the affect of asymmetry of the $\lambda$-anomaly to the quasi-2D character of the magnetic structure, i.e., to the fact that the buckled (100) layers of strongly coupled Co$^{2+}$ magnetic moments are not isolated but weakly coupled mutually. With increasing $B$, $T_N$ decreases parabolically, as illustrates the inset to Fig.\ \ref{fig2}b, and the $\lambda$-anomaly decreases, but the transition remains sharp. \ Such a behaviour is characteristic of a 2D antiferromagnetic Ising system.\cite{fisher}

An additional anomaly appears in $B=8$\ T at 8.8\ K, Figs.\ \ref{fig2}b, \ref{fig3}b, and \ref{fig4}a.\ For $B=9$ T, it becomes more pronounced and shifts to 9.2 K. Near the anomaly, a hysteresis between the curves measured on heating and on cooling appears and the anomaly measured on cooling is smaller, Fig.\ \ref{fig3}b. Since these two effects are the basic characteristics of first-order transitions\cite{szewc71, szewc72} (the first one is related to overheating and overcooling phenomena and the second one is inherent in the relaxation method of measurement), we interpret the anomaly as the indication of occurrence of a first order phase transition. It can be supposed that the tendency observed on increasing the magnetic field from 8 to 9 T will be preserved and the anomaly will increase and shift to higher temperatures with further increase of the magnetic field.

To estimate a change of magnetic entropy related to this transition, we assumed that the magnon contribution, $C_\text{ma}$, being the only constituent of $C_m$ apart from the transition, can be described in frames of the model developed for anisotropic antiferromagnets.\cite{akhi} For the cases of low and high temperatures, that model predicts, respectively:

\begin{equation}
C_\text{ma}=a_1 \frac{1}{\sqrt{T}}\exp{\left (-\frac {a_2}{T} \right )}   \text{~~for~~}  \frac{\mu_B}{k_B} B_a > \frac{\mu_B}{k_B} B > T,
\label{eq3}
\end{equation}
\begin{equation}
C_\text{ma}=a_3T^3     \text{~~~~for~~~~}  T_N \gg T \gg  \frac{\mu_B}{k_B} B_a > \frac{\mu_B}{k_B} B,
\label{eq4}
\end{equation}
where $a_1=a_0(B_a-B)^2$, $a_2=b_0(B_a-B)$, $a_0$, $b_0$, and $a_3$ are constants, $B_a$ is a parameter of the order of anisotropy and exchange fields, $\mu_B$ is Bohr magneton, and $k_B$ is Boltzmann's constant. A good description was achieved up to 14 K, Fig.\ \ref{fig4}a. For $B < 5$ T, the experimental dependences can be fitted with Eq.\ (\ref{eq3}), whereas for larger fields a crossover between the behaviors given by Eqs.\ (\ref{eq3}) and (\ref{eq4}) occurs at $\sim$\,9.5 K. To get a satisfactory description above the crossover temperature for $B$\,=\,7, 8, and 9 T, it is necessary to add to Eq.\ (\ref{eq4}) a constant term, respectively, of 0.31, 0.41, and 0.53 J/(mole K). Tentatively, this can be ascribed to the fact that the model of noninteracting magnons,\cite{akhi} within which Eqs. (\ref{eq3}) and (\ref{eq4}) were derived, is inadequate close to $T_N$ (which falls down from $\sim$\,22 K for $B$\,=\,0 to $\sim$\,18 K for $B$\,=\,9 T). A steep changes of the calculated magnon contributions appearing at the transition point for $B \geqslant 8$ T, Fig.\ \ref{fig4}a, suggest that the transition is related to a change in the stiffness of the magnon system, i.e. to a change of anisotropy or exchange interactions. After subtracting the calculated magnon contributions from $C_m$, Fig.\ \ref{fig4}a, the entropy change associated with the first-order transition, $\Delta S$, was calculated using the formula: $\Delta S= \int_{T_1}^{T_2} \left [ \left ( C_m - C_\text{ma} \right ) /T \right ] dT $ (the temperatures $T_1$ and $T_2$ must be chosen sufficiently far from the transition temperature, i.e., at points at which $C_m = C_\text{ma}$). It was found $\Delta S /(k_BN_A) = 0.010$, which is a very small value (e.g., the entropy change related to disappearance of a long-range order in a system of 1/2 spins is equal to $\text{ln(2)} \approx \text{0.693}$).

Based on Ref.\ \onlinecite{wieg}, where the magnetoelectric effect was shown to vanish in $B\sim$\,12 T, and on Ref.\ \onlinecite{khar10}, where destruction of the antiferromagnetic ordering in the magnetic field parallel to the \textit{b} axis was shown to occur via a series of spin-flip transitions, starting from $B\sim$\,12 T, we claim that the magnetic field of 9 T: (i) can influence the distribution of electric charges and, hence, the magnetocrystalline anisotropy of LiCoPO$_4$, (ii) is too weak to induce a spin-flip transition. In order to verify these claims and to elucidate the physical nature of the discovered transition, supplementary magnetization and magnetic torque measurements have been performed.

\begin{figure}
\includegraphics[]{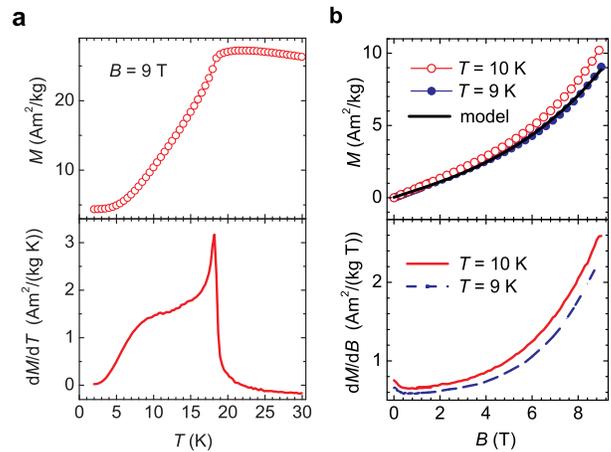}
\caption{\label{fig5}(Color online).
Magnetization of LiCoPO$_4$ along the \textit{b} axis, measured in the magnetic field applied along the \textit{b} axis. (a) Magnetization in $B=9$ T (and its derivative with respect to $T$) as a function of temperature. (b) Magnetizations at $T=9$ and 10 K (and their derivatives with respect to $B$) as a function of magnetic field. Solid line superimposed on the experimental dependence for 9 K was calculated within the two-sublattice model proposed.}
\end{figure}

The magnetization studies, Fig.\ \ref{fig5}, confirmed that this is not a spin-flip transition, because neither in the temperature and field dependences of magnetization nor in their first derivatives any anomaly occurs at the transition point.

\begin{table*}
\caption{\label{tab2} Parameters, which fit the best the measured dependences of torque on $\theta $ and of magnetization along the \textit{b} axis on $B$. Numbers in parentheses give uncertainty of the last digit (or of the two last digits) of the parameters, e.g., 21.72(4) means $21.72\pm 0.04$. Parameters that changed as the result of the phase transition are given in bold. $T_0$ is a small constant value (approximately, two orders of magnitude smaller than the maximum torque values measured) that must be added to the theoretical dependence of torque on $\theta$, (\ref{app19}), to offset a background, inherent in the measurement technique applied.}
\begin{ruledtabular}
\begin{tabular}{ccccccccccc}
$T$ & $B$  & $K_1$ & $B_m$ & $\chi_{1b}$ &  $\chi_{2b}$ & $\chi_{3b}$ &  $\chi_{1c}$ &  $\chi_{2c}$ & $\chi_{3c}$ & $T_0$ \\
(K) & (T) & $(10^6$J/m$^3)$ & (T) & $(10^3 \text{A/(m\,T)})$ & $(10^3 \rm{A/(m\,T^2)})$ & $(\rm{A/(m\,T^3)})$ &  $(10^3\text{A/(m\,T)})$ & $(10^3 \rm{A/(m\,T^2)})$ & $(\rm{A/(m\,T^3)})$ & $(10^{-7}$Nm $)$\\
\hline
9&6&4.3(3)& 20.0(2) & 28.01(3) & -1.304(2) & 21.72(4) &  8.7(3) & -1.2(1) & 50(30) & -2(1) \\
9&7&4.3(3)& 20.0(2) & 28.01(3) & -1.304(2) & 21.72(4) &  9.4(2) & -1.25(5) & 50(20) & -2(1)\\
9&8&4.3(3)& 20.0(2) & 28.01(3) & -1.304(2) & 21.72(4) &  9.9(1) & -1.26(2) & 50(10) & -2(1) \\
9&9&4.3(3)& 20.0(2) & 28.01(3) & -1.304(2) & 21.72(4) &  10.0(1)& -1.16(2) & 50(5) &  -2(1) \\
\hline
12&6&4.3(3)& 17.55(5) & 29.90(2) & -1.520(1) & 28.87(8) & 5.8(1) & -0.33(3) & 140(20) & -2(1) \\
12&9& \textbf{11.5}(1.1) & 17.55(5) & 29.90(2) & -1.520(1) &  28.87(8) & \textbf{6.36}(2) & \textbf{-0.113}(3) & \textbf{150}(5) & -3(1)\\
\end{tabular}
\end{ruledtabular}
\end{table*}

The magnetic torque has been measured at $T$\,=\,9 and 12 K, for the magnetic field of different value, rotating within the $b$-$c$ plane, Figs.\ \ref{fig4}b and \ref{fig4}d. The angle $\theta$ determining the orientation of the field $B$ within the $b$-$c$ plane was counted from the \textit{b} axis. The torque measurements showed unequivocally, Fig.\ \ref{fig4}d, that the transition is related to a change in the magnetic anisotropy. This follows from the fact that at $T=9$ K, i.e., below the transition point, the positions of the torque maxima evolve monotonically with increasing \textit{B} (up to the highest \textit{B} value of 9 T) along the direction indicated in Fig.\ \ref{fig4}d by a blue oblique arrow, while at $T=12$ K, i.e., above the transition point, we observe a steep, qualitative change of the torque behavior. That means at $T=12$ K, for $B=6$ T, the maxima fit into the tendency observed at $T=9$ K, whereas for $B=9$ T, the maxima shift to the opposite direction than at $T = 9$ K. This is illustrated in Fig.\ \ref{fig4}d by a red oblique arrow.

To describe this effect theoretically, we propose a simplified model, Fig.\ \ref{fig4}c, details of which are presented in the appendix, based on the following assumptions:
\begin{enumerate}
\item[(i)] As it was suggested in Ref. \onlinecite{khar28}, modulated in space, perpendicular to the \textit{b} axis, nonzero components of magnetic moments exist in the magnetic structure of LiCoPO$_4$.
\item[(ii)] The two-sublattice model, in which Co2 and Co3 ions form the one sublattice, denoted as 1, and Co1 and Co4 ions form the other one, denoted as 2, can be applied.
\item[(iii)] The modulated, perpendicular to the \textit{b} axis components average out to zero and only the net magnetizations of both sublattices are essential, Fig.\ \ref{fig4}c.
\item[(iv)] The deflection of the net magnetizations of both sublattices by 4.6$^{\circ}$ from the \textit{b} axis and the possibility of existence of domains, in which the sign of this deflection is different, can be neglected and it can be assumed that the net magnetizations are directed along the \textit{b} axis.
\item[(iv)] The sublattice magnetized ``along" the field (denoted as $\mathbf{M}_1$ in Fig.\ \ref{fig4}c) is rigid, has a well defined magnetization modulus, and its magnetic anisotropy can be analyzed by using the anisotropy constant $K_1$.
\item[(v)] The sublattice magnetized ``against" the external field (denoted as $\mathbf{M}_2$) is ``weak" and behaves as an anisotropic paramagnet located within an effective field, $\mathbf{B}_{\text{eff}}$, composed of the external field, $\mathbf{B}$, and the exchange (i.e. molecular) field, $\mathbf{B_\emph{m}}$, produced by the ``rigid" sublattice.
\item[(vi)] The magnetization of the ``weak" sublattice is equal to:
$M_{2\sigma} = \chi_{1\sigma} B_{\text{eff}\,\sigma} + \chi_{2\sigma} \text{sign}(B_{\text{eff}\, \sigma}) B_{\text{eff}\,\sigma}^2+ \chi_{3\sigma} B_{\text{eff}\,\sigma}^3$, where $ \sigma $ identifies the components along the $b$ and $c$ axes.
\item[(vii)] For $-90^{\circ} < \theta < 90^{\circ}$, the sublattice 1 is the rigid one and the sublattice 2 is the susceptible one. For $90^{\circ} < \theta < 270^{\circ}$, both sublattices exchange their behavior, that means the sublattice 1 becomes the susceptible one, while the sublattice 2 becomes the rigid one.
\item[(viii)] For $\theta = \pm 90^{\circ}$ both sublattices are indistinguishable, therefore the torque goes through zero at these angles.
\end{enumerate}

By taking the $B_m$, $\chi_{1b}$, $\chi_{2b}$, and $\chi_{3b}$ values, which fit the best the dependences of net magnetization on $B$ applied along the \textit{b} axis (measured in  Ref.\ \onlinecite{khar28} and in the present work, Fig.\ \ref{fig5}b), and by treating $K_1$,  $\chi_{1c}$, $\chi_{2c}$, and $\chi_{3c}$ as fitted parameters, a good agreement between the measured and the calculated angle dependences of torque was achieved (Fig.\ \ref{fig4}b). The values of the parameters fitting the experimental data the best are given in Table \ref{tab2}. These values imply that at the transition point the sublattice magnetized ``along" the field becomes ``harder" ($K_1$ grows), whereas the other sublattice becomes ``weaker" (its total susceptibility along the \textit{b} axis remains unchanged, whereas the susceptibility along the \textit{c} axis grows). Nevertheless, the \textit{b} axis remains the easy magnetization direction (the sign of $K_1$, characterizing the anisotropy of the sublattice 1, does not change and the total susceptibility of the sublattice 2 remains larger along the \textit{b} axis than along the \textit{c} axis). This fact explains why no anomalies are observed at the transition point on temperature and field dependences of magnetization measured in the magnetic field applied along the \textit{b} axis, Fig.\ \ref{fig5}. The uncertainty of each parameter given in Table \ref{tab2} was estimated by keeping all other parameters fixed and checking that no noticeable change of the theoretical curve appears for the values of the examined parameter lying within the uncertainty range, whereas for the values beyond that range, fit quality deteriorates evidently. It should be mentioned that the proposed method of analysis, in which 9 fitted parameters are involved, should be treated rather as a qualitative method of elucidating the physical processes occurring in the sample, not as an accurate method for determining physical parameters, e.g. $K_1$.

\section*{conclusions}

In conclusion, it was shown that in LiCoPO$_4$, the second order phase transition from the paramagnetic to the weakly ferromagnetic phase is accompanied by a $\lambda$-shaped anomaly of specific heat, which can be described as the logarithmic divergence (\ref{eq2}), characteristic of a 2D Ising system. The deviation from the purely 2D behaviour was ascribed to the quasi-2D character of the magnetic structure. The first-order phase transition induced by an external magnetic field $B \geqslant 8$ T parallel to the \textit{b} axis, appearing at $\sim$\,9 K, was discovered and shown to be related to the change of magnetocrystalline anisotropy.

\begin{acknowledgments}
Support of Swiss NSF for crystal growth (prior to 1996) is gratefully acknowledged. This work was partly supported by the Polish Ministry of Science and Higher Education from funds for science for 2008-2011 years, as a research project (2047/B/H03/2008/34), and by the European Union, within the European Regional Development Fund, through the Innovative Economy grant (POIG.01.01.02-00-108/09).
\end{acknowledgments}

\appendix*
\section{}

\begin{figure}
\includegraphics[scale=0.7]{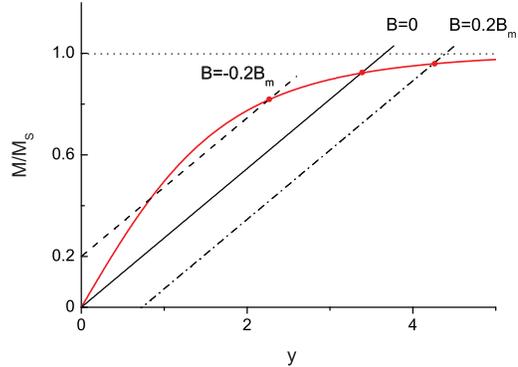}
\caption{\label{fig6}(Color online) Molecular field approximation for a standard ferromagnet. Red curve presents the Brillouin function. Straight solid line presents the case of zero external field. The dash-dot and dashed lines show, respectively, the cases of the external field applied along and against the molecular field.
}
\end{figure}

We assumed that the simplest description of the LiCoPO$_4$ antiferromagnet can be based on the two-sublattice approximation, in which Co2 and Co3 ions form the one sublattice, denoted as 1, and Co1 and Co4 ions form the other one, denoted as 2 (Figs.\ \ref{fig1} and \ref{fig4}c). Then, we can apply a molecular field approximation, which must be modified in such a way that the presence of an extremely weak, net spontaneous magnetization, $M_{\text{sp}}^{\text{ex}}$, found experimentally,\cite{khar28, khar27} will be mimicked. An approach to the latter effect can be based on considerations concerning a much simpler case of a ferromagnet. It is known that within the molecular field approximation, the magnetization of a ferromagnet can be determined graphically as the intersection point of the Brillouin function:
\begin{equation}
B_S(y)=\frac{2S+1}{2S}\coth \left(\frac{2S+1}{2S}y\right)- \frac{1}{2S} \coth \left(\frac{1}{2S}y \right),
\label{app1}
\end{equation}
and of the linear relation between the $y$ parameter and the magnetization:
\begin{equation}
\frac{M}{M_S}=\frac{k_B T}{g\mu_B S B_m }y - \frac{1}{B_m}B,
\label{app2}
\end{equation}
where $B_m$, $M_S$, $g$, and $S$ denote, respectively, molecular field, saturation magnetization, $g$-factor and spin of the magnetic ion.
As shown in Fig.\ \ref{fig6}, for $T=0.5 \; T_C$, in zero magnetic field the intersection point is located in the plateau of the Brillouin function and the spontaneous magnetization reaches nearly the saturation value $M_S$. Then, if the external field $B=0.2 \; B_m$ is applied along $B_m$, the magnetization changes only slightly, whereas if the same field is applied against $B_m$, the intersection point shifts to the region of noticeably smaller magnetization values, where also the curvature of the Brillouin function and the susceptibility are larger.

Since for LiCoPO$_4$, the temperature of the first-order transition is  $\sim \; 0.5 \; T_N \sim\;10$ K, $B_m$ (estimated based on the $T_N$ value) is $\sim$18 T, and the applied fields ranging from 6 to 9 T are of the order from 0.3 to 0.5 $B_m$, we can expect a similar behavior for this more complex case of weakly ferromagnetic antiferromagnet.

\begin{figure}
\includegraphics[scale=0.25]{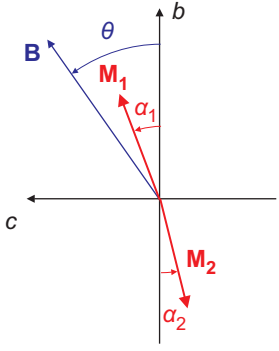}
\caption{\label{fig7}(Color online) Coordinate system and the symbols used.
}
\end{figure}

Thus, we assumed that the sublattice 1, magnetized along the field, Figs.\ \ref{fig4}c and \ref{fig7}, is rigid, with well defined modulus of its magnetization, $M_0 = 2 m_{\text{Co}}/V_{\text {uc}}$, where $m_{\text{Co}} = g\mu_B S = 3.26 \mu_B$ is the magnetic moment of one Co$^{+2}$ ion ($S=3/2$) and $V_{\text {uc}}$ is the volume of the orthorhombic unit cell (containing 4 formula units). As the result, magnetic anisotropy of this sublattice can be described by using the anisotropy constant $K_1$. On the contrary, the sublattice 2, magnetized against the field, is susceptible and the modulus of its magnetization is a function of value and direction of the field $B$. Thus, the formalism involving anisotropy constants is inapplicable and we can describe a magnetic anisotropy of the sublattice 2 by introducing different magnetic susceptibilities along different crystallographic directions.

Additionally, the following experimental facts \cite{khar28, khar27} must be taken into account:

\noindent (i) The dependence of spontaneous magnetization, $M_{\text{sp}}^{\text{ex}}$, on temperature has a form found in Ref.\onlinecite{khar28}:
\begin{equation}
M_{\text{sp}}^{\text{ex}}(T) = \text{N}(0.122 - 6.5 \times 10^{-4}(T-10.4)^2),
\label{app3}
\end{equation}
where N is a coefficient needed to convert the value expressed in G to desired units.

\noindent (ii) The dependence of magnetization on the field applied along the \textit{b} axis contains terms linear and cubic in $B$, as it was found in Ref.\onlinecite{khar27}:
\begin{equation}
M(B)=M_{\text{sp}}^{\text{ex}}(T) + \chi_{1}^{\text{ex}}B + \chi_3^{\text{ex}} B^3.
\label{app4}
\end{equation}

(Coming beyond the molecular field approximation we can say that the real LiCoPO$_4$ magnetic structure can have a form of a very weakly spread out "fan",\cite{khar2000} Fig.\ \ref{fig4}c. Then, we can imagine that the molecular field picture described above is a simplification of the fact that the "fan" of the sublattice magnetized along the filed folds slightly, whereas the fan of the sublattice magnetized against the field spreads out considerably under influence of the magnetic field.)

Besides we assumed that the deflection of the magnetization by 4.6$^{\circ}$ away from the \textit{b} axis and the possibility of existence of domains differing in sign of this deflection, i.e., $+4.6^{\circ}$ or $-4.6^{\circ}$, has no noticeable effect on the angle dependences of torque and on the net magnetization value along the \textit{b} axis and can be neglected. Thus, in our considerations was assumed that the \textit{b} axis is the easy magnetization direction for both sublattices. Validity of this assumption has been verified by direct calculations.

Under the assumptions given above, for $-90^{\circ} \leqslant \theta \leqslant 90^{\circ}$ the magnetizations of both sublattices are given by the expressions:
\begin{equation}
M_{1b} = M_0 \cos \alpha_1, \qquad  M_{1c} = M_0 \sin \alpha_1,
\label{app5}
\end{equation}
\begin{equation}
M_{2\sigma} = \chi_{1\sigma} B_{\text{eff}\,\sigma} + \chi_{2\sigma} \text{sign}(B_{\text{eff}\, \sigma}) B_{\text{eff}\,\sigma}^2+ \chi_{3\sigma} B_{\text{eff}\,\sigma}^3
\label{app6}
\end{equation}
where $\sigma$ denotes $b$ and $c$. The effective field, $\mathbf{B}_{\text{eff}}$, acting on the sublattice 2 consists of the applied, $\mathbf{B}$, and the molecular, $\mathbf{B}_m$, field and is equal to:
\begin{eqnarray}
B_{\text{eff}\,b}=B \cos \theta - B_m \cos \alpha_1 \nonumber \\
B_{\text{eff}\,c}=B \sin \theta - B_m \sin \alpha_1.
\label{app7}
\end{eqnarray}
Then, the free energy of the system is given by the expression:
\begin{equation}
\begin{split}
&F(T,B,\alpha_1,\theta)= K_1\sin^2\alpha_1 - M_0 B\cos (\theta - \alpha_1) \\
& -\sum_{\sigma = b,c} \left(\frac{1}{2} \chi_{1\sigma} B_{\text{eff}\,\sigma}^2+\frac{1}{3} \chi_{2\sigma}\lvert B_{\text{eff}\,\sigma}\rvert^ 3 + \frac{1}{4}\chi_{3\sigma} B_{\text{eff}\,\sigma}^4 \right).
\label{app8}
\end{split}
\end{equation}

To assure consistency with the experimental results of magnetization measurements, we assume that for $\theta = 0$ and $B$ smaller than the spin-flip field (as it is in the considered case), also $\alpha_1 = 0$ and $d \alpha_1 / dB = 0$. Then, the theoretical resultant magnetization of the sample, calculated according to the formula:
\begin{equation}
M (\theta =0, B)=-\frac{\partial F}{\partial B} = M_0+M_{2b},
\label{app9}
\end{equation}
should be equal to the experimental one, given by Eq.\ (\ref{app4}). By comparing coefficients at different powers of $B$ (in particular, the coefficient at $B^2$ should be equal to 0) we receive the following relations:
\begin{equation}
\begin{split}
\chi_{1b}=&\chi_1^{\text{ex}}+3\chi_3^{\text{ex}}B_m^2, \quad \chi_{2b}=-3\chi_3^{\text{ex}}B_m, \\ \chi_{3b}= &\chi_3^{\text{ex}}.
\end{split}
\label{app15}
\end{equation}
By substituting $B=0$ into (\ref{app9}), we receive the equation:
\begin{equation}
M_0 - \chi_1^{\text{ex}}B_m - \chi_3^{\text{ex}}B_m^3 = M_{\text{sp}}^{\text{ex}}(T).
\label{app16}
\end{equation}

Next, for $T = 9$ and 12 K, the $M_{\text{sp}}^{\text{ex}}(T)$ parameter was calculated according to Eq. (\ref{app3}) and the coefficients $\chi_1^{\text{ex}}$ and $\chi_3^{\text{ex}}$ were determined by fitting the function (\ref{app4}) to the $M(B)$ dependences (since we had access to the $M(B)$ dependences measured for $T = 9$ and 10 K only, the $\chi_1^{\text{ex}}$ and $\chi_3^{\text{ex}}$ coefficients for $T = 12$ K were determined by extrapolating the ones found for $T = 9$ and 10 K). Then, knowing these parameters, $B_m$ was determined by solving Eq. (\ref{app16}) and the susceptibilities $\chi_{ib}$, for \textit{i} = 1, 2, and 3, were calculated using Eq.\ (\ref{app15}). (We found $B_m= 20$ T for $T=9$ K and $B_m= 17.55$ T for $T=12$ K. At the first glance it seems strange that $B_m$
varies so considerably, but this can be attributed to the unusual parabolic dependence of the spontaneous magnetization on $T$ (\ref{app3})).

In order to determine angle dependences of the magnetic torque for $\mathbf{B}$ rotating within the $b-c$ plane, for the case $-90^{\circ} \leqslant \theta \leqslant 90^{\circ}$, the $\alpha_1$ parameter was determined for each $\theta$ value by solving (numerically) the entangled equation:
\begin{equation}
\begin{split}
&\frac{\partial F}{\partial \alpha_1}=\left(\frac{2K_1}{M_0}\right)M_0 \sin\alpha_1 \cos \alpha_1 - M_0 B\sin(\theta - \alpha_1)  \\
&-\bigl(\chi_{1b} + \text{sign}(B_{\text{eff}\,b}) \chi_{2b} B_{\text{eff}\,b} + \chi_{3b}B_{\text{eff}\,b}^2 \bigr) B_{\text{eff}\,b}B_m \sin\alpha_1 \\
& + \bigl(\chi_{1c}+ \text{sign}(B_{\text{eff}\,c})\chi_{2c}B_{\text{eff}\,c} + \chi_{3c}B_{\text{eff}\,c}^2 \bigr)B_{\text{eff}\,c}B_m \cos\alpha_1\\
&=0.
\label{app18}
\end{split}
\end{equation}
Next, $M_{1\sigma}$ and $M_{2\sigma}$ values were calculated by using the formulae (\ref{app5}) - (\ref{app7}) and the magnetic torque acting on the whole sample was calculated according to the formula:
\begin{equation}
\begin{split}
&T_a= V \left(\mathbf{M}\times \mathbf{B}\right)_a= \frac{m}{m_m} \frac{\text{N}_\text{A}}{4} V_{\text{uc}} \Bigl[(M_{1b}+M_{2b})B_c \\
&- (M_{1c}+M_{2c})B_b \Bigr]= \frac{m}{m_m} \frac{\text{N}_\text{A}}{4} V_{\text{uc}}B \Bigl[M_0\sin(\theta-\alpha_1) \\
&+\bigl(\chi_{1b}+\text{sign}(B_{\text{eff}\,b})\chi_{2b}B_{\text{eff}\,b}+\chi_{3b}B_{\text{eff}\,b}^2 \bigr)B_{\text{eff}\,b}\sin\theta \\
&-\bigl(\chi_{1c}+\text{sign}(B_{\text{eff}\,c})\chi_{2c}B_{\text{eff}\,c}+\chi_{3c}B_{\text{eff}\,c}^2 \bigr)B_{\text{eff}\,c}\cos\theta   \Bigr],
\label{app19}
\end{split}
\end{equation}
where $V$, $m$, $m_m$, and N$_\text{A}$ denote, respectively, the volume of the sample, the mass of the sample, the LiCoPO$_4$ molar mass, and the Avogadro number.
The $\alpha_2$ and $M_2$ values were determined by using the formulae:
\begin{equation}
\alpha_2=\arctan\left(\frac{M_{2c}}{M_{2b}}\right), \quad M_2=\sqrt{M_{2b}^2+M_{2c}^2}.
\label{app20}
\end{equation}

For the case $90^{\circ} < \theta < 270^{\circ}$, both sublattices exchange their behavior, that means the sublattice 2 becomes rigid, whereas the sublattice 1 becomes susceptible. Thus, the appropriate free energy and torque values were calculated by using Eqs. (\ref{app7}), (\ref{app8}), (\ref{app18}), and (\ref{app19}), in which $\theta$ was replaced with $\theta' = \theta - 180^{\circ}$ and the $\alpha_1$ parameter determined by solving Eq. (\ref{app18}) was treated as $\alpha_2$.

It should be mentioned that for the special values $\theta = \pm 90^{\circ}$ both sublattices are indistinguishable and the torque is equal to zero.

As the result of applying the procedure described above, in which $K_1$,  $\chi_{1c}$, $\chi_{2c}$, and $\chi_{3c}$ were treated as fitted parameters, the theoretical dependences of $T_a$ on $\theta$, plotted in Fig. \ref{fig4}b with (black) solid lines superimposed on the experimental data, have been obtained.
\vspace{1.5cm}


%

\end{document}